\documentclass[twocolumn, floatfix, showpacs]{revtex4}

\usepackage[final]{epsfig}
\usepackage{amsmath}
\usepackage{parskip}

\begin{document}

\title{Thermal photon spectra and HBT correlations at full RHIC energy}

\author{Thorsten Renk}

\preprint{DUKE-TH-04-263}

\affiliation{Department of Physics, Duke University, PO Box 90305,  Durham, NC 27708 , USA}

\begin{abstract}
Using a model for the evolution of the fireball created in
Au-Au collisions at full RHIC energy which reproduces the hadronic single particle spectra and two-particle
correlations we calculate thermal photon
emission from the hot partonic and hadronic matter.  We present predictions
for both the emitted photon transverse momentum spectrum and 
Hanbury-Brown Twiss (HBT) correlations. 
Surprisingly, we
find that due to the strong transverse flow in the model photon emission from
hadronic matter is sizeable in the momentum range expected to be dominated
by thermal emission and that consequently the HBT radius parameters do not
reflect exclusively properties of the partonic evolution phase. We compare to
another model calculation where pre-equilibrium photon emission is
dominant to illustrate the differences of both approaches.

\end{abstract}

\maketitle

\section{Introduction}
\label{sec_introduction}

In the hot and dense system created in a heavy-ion collision, the relevant momentum scales
for scattering processes drop as a function of proper time as a large number of 
particles is created and the system
evolves towards equilibrium. Assuming that an expanding thermalized system is created after 
some initial timescale $\tau_0$, the relevant momentum scales inside the thermalized matter 
are then set by the temperature, which in turn drops as the fireball volume expands
with time. Thus, by selecting 
a certain window in momentum for scattering processes, one  is preferentially sensitive 
to a certain proper time window of the evolution. If this window is chosen such that it is 
below the scales associated with non-equilibrium initial scattering processes, one is, 
for $\langle p\rangle \approx 3T_i$ with $T_i$ the initial temperature of the system, 
in principle sensitive to the initial hot and dense phase where hadronic matter presumably undergoes 
a phase transition to the quark-gluon plasma (QGP) as seen in lattice QCD simulations (see e.g. \cite{Lattice}).

Since electromagnetic probes (such as real photons and dileptons) can escape from the system at all times due to the relative smallness of the electromagnetic coupling constant, a measurement of hard real 
photon emission can be expected to provide information about the early phases of fireball
evolution and ultimately a window to observe the QGP directly \cite{Feinberg, Shuryak}. 
Several such investigations have been carried out for the CERN SPS (see e.g. \cite{Huovinen, Srivastava})
In a schematic way this was addressed also in \cite{Photons} where data measured by
the WA98 data \cite{PhotonData} could indeed be used to find constraints for the
equilibration time. In addition, the measured photon spectrum allowed to distinguish between
a boost-invariant longitudinal expansion and a scenario involving initial compression and
re-expansion, the latter being favoured by other observables as well \cite{Comprehensive}.

In a recent paper, an approach has been suggested aiming at determining the relevant scales of the
spacetime evolution of a fireball created in an Au-Au collision at full RHIC energy
\cite{RHIC-Model}. It is the aim of this paper to explore what amount of
thermal photon emission is predicted from the resulting evolution and to what degree 
a precise measurement of the photon spectrum is able to further constrain the evolution dynamics of the
system. In addition to photon transverse momentum spectra, we also discuss predictions
for Hanbury-Brown Twiss (HBT) correlation measurements of hard photons and their
capability to provide additional information about the early stages of the system.

\section{The model}

In this section, we present a
model framework for the fireball expansion which in essence is a parametrization
inspired by a hydrodynamical evolution of the collision system. This model has been
shown to give a good description of both single particle spectra and two particle
correlations at RHIC simultaneously for a breakup temperature well below the
phase transition temperature. It is described
in greater detail in \cite{RHIC-Model}, here we only repeat the essential facts:

For the entropy density at a
given proper time we make the ansatz 
\begin{equation}
s(\tau, \eta_s, r) = N R(r,\tau) \cdot H(\eta_s, \tau)
\end{equation}
with $\tau $ the proper time measured in a frame co-moving with a given volume element, 
$\eta_s = \frac{1}{2}\ln (\frac{t+z}{t-z})$ the spacetime
rapidity and $R(r, \tau), H(\eta_s, \tau)$ two functions describing the shape of the distribution
and $N$ a normalization factor. We use Woods-Saxon distributions 
\begin{equation}
\begin{split}
&R(r, \tau) = 1/\left(1 + \exp\left[\frac{r - R_c(\tau)}{d_{\text{ws}}}\right]\right)
\\ & 
H(\eta_s, \tau) = 1/\left(1 + \exp\left[\frac{\eta_s - H_c(\tau)}{\eta_{\text{ws}}}\right]\right).
\end{split}
\end{equation}
for the shapes. Thus, the ingredients of the model are the skin thickness 
parameters $d_{\text{ws}}$ and $\eta_{\text{ws}}$
and the para\-me\-tri\-zations of the expansion of 
the spatial extensions $R_c(\tau), H_c(\tau)$ 
as a function of proper time. From the distribution of entropy density, the thermodynamics can be inferred
via the EoS and particle emission is then calculated using the Cooper-Frye formula.
For simplicity, we apply the model at the moment only to central and almost central collisions and
assume that the flow is built up by a constant acceleration $a_\perp$, hence 
$R_c(\tau) = R_c^0 + \frac{a_\perp}{2} \tau^2$
with $R_c^0$ an initial radial extension as found in overlap calculations
In \cite{RHIC-Model}, the model parameters have been adjusted such that the model gives a good
description of the data.

\section{The photon emission rate}

The complete calculation of the photon emission rate from a partonic medium 
to order $\alpha_s$ has been a very involved task 
\cite{2-2-Kapusta,2-2-Baier,Aurenche1, Aurenche2, Aurenche3,Complete1, Complete2}.
The relevant processes include $q\overline{q}$ annihilation, QCD compton 
scattering, bremsstrahlung and $q\overline{q}$ annihilation with subsequent
scattering.
For the present calculation, we use the parametrization of the rate
given in \cite{Complete2}.

Vector mesons play an important role for the emission
of photons from a hot hadronic gas. The first calculation
of such processes has been performed in \cite{Kapusta-Eff}
in the framework of an effective Lagrangian.
It has been found that the dominant processes
are pion annihilation, $\pi^+\pi^- \rightarrow \rho \gamma$,
'Compton scattering', $\pi^\pm \rho \rightarrow \pi^\pm \gamma$
and $\rho$ decay, $\rho \rightarrow \pi^+\pi^- \gamma$.

Several more refined approaches have been made since then
(for an overview, see \cite{PhotonReview}).
In the following, we will use a parametrization of the
rate from a hot hadronic gas taken from \cite{HHG}.

\subsection{The photon spectrum}

The spectrum of emitted photons can be found by folding the rate with the
fireball evolution. In order to account for flow, the energy of a photon emitted
with momentum $k^\mu =(k_t, {\bf k_t}, 0)$ has to be evaluated in the local rest
frame of matter, using $E=k^\mu u_\mu$ with $u_\mu(\eta_s, r, \tau)$
the local flow profile. Following the results in \cite{RHIC-Model} we assume 
$\eta = const. \cdot \eta_s$ and $\eta_\perp = const. \cdot r$ with $\eta_\perp$ the transverse
rapidity in the following. The distribution of entropy density is manifest
in the dependence of the temperature $T = T(\eta_s, r, \tau)$ on the spacetime
position. In order to account for the breakup of the system once a temperature
$T_F$ is reached, a factor $\theta(T-T_F)$ has to be included into the folding
integral. 

In order to be able to compare with SPS data and our previous model calculations, we present the 
differential emission spectrum into the midrapidity slice $y=0$.

\section{Photon spectra}

\begin{figure*}[htb]
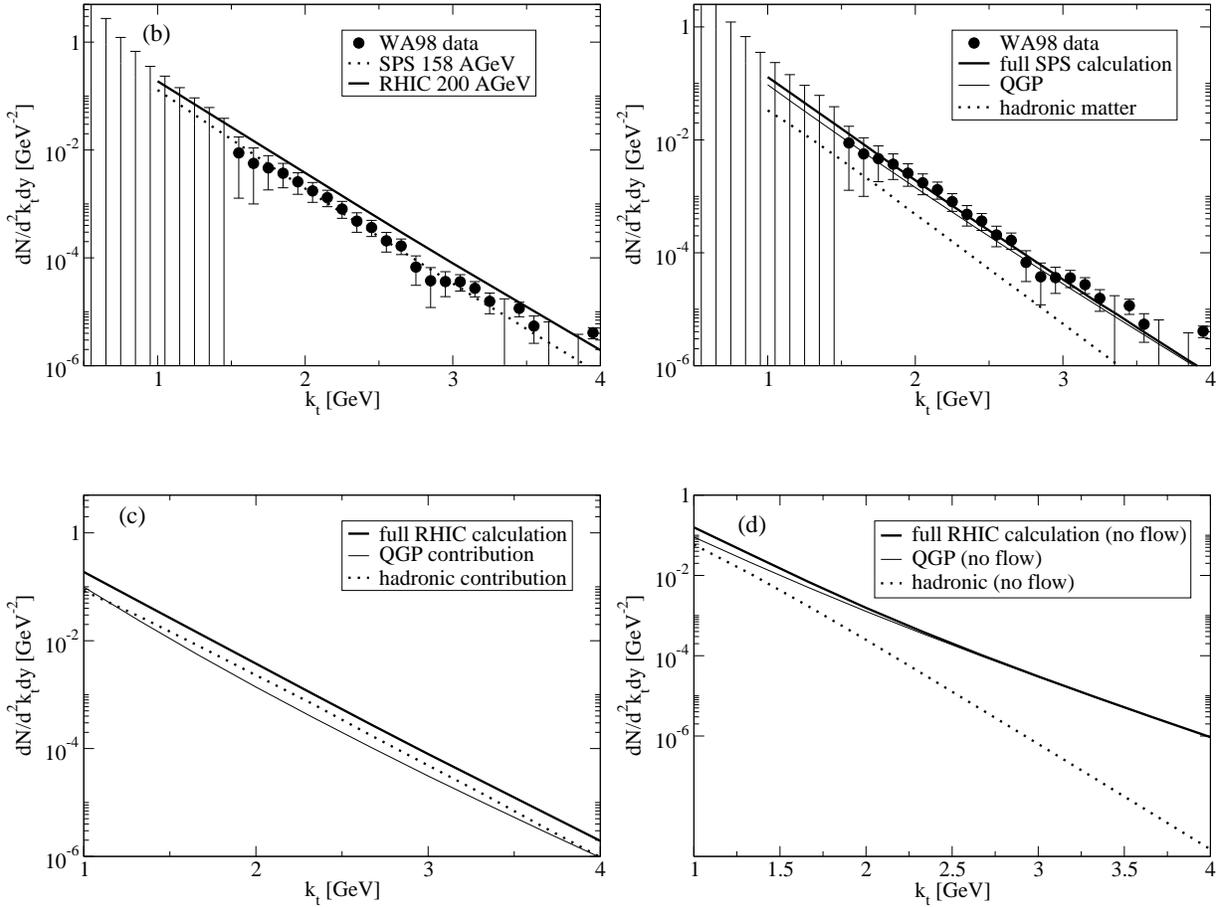

\epsfig{file=photon_rhic_new.eps, width=8cm}
\epsfig{file=photon_data.eps, width=8cm}

\vspace*{0.9cm}

\epsfig{file=photon_rhic_phases_new.eps, width=8cm}
\epsfig{file=photon_rhic_noflow_new.eps, width=8cm}
\caption{\label{F-1}Upper left: The direct photon spectrum measured at the CERN
SPS \cite{PhotonData} as compared with the model calculations for SPS (dotted) and
RHIC (solid). Upper right: The measured spectrum at SPS, shown with the model calculation for
SPS (solid) and the contributions from QGP (thin) and hadron gas (dotted). Lower left: The photon
spectrum in the model calculation for RHIC (solid) and the contributions from QGP (thin) and
hadron gas (dotted). Lower right: As before, but without taking transverse flow into account
when evaluating the photon emission from a given volume element.}
\end{figure*}

The resulting photon spectra are shown in Fig.~\ref{F-1}. In (a), we compare the calculation for
RHIC with a model calculation for SPS done in the framework outlined above (essentially this
reproduces the results obtained in \cite{Photons} for SPS in a similar, but slightly less
sophisticated framework) and the data obtained for SPS 158 AGeV 10\% central Pb-Pb collisions
\cite{PhotonData}.

As expected, the RHIC photon spectrum is characterized by a larger absolute
normalization (reflecting a larger four-volume of radiating matter)
and a less steep falloff, reflecting a higher temperature of the emission
region and/or a larger amount of transverse flow. Roughly, we find a factor
two more photons at 2 GeV at RHIC than at SPS.

In  order to investigate this ambiguity between flow and
temperature in more detail, we compare in (b) and (c)
the amount of photon radiation coming from the QGP phase
and the hadronic phase, respectively. Surprisingly, while
at SPS photon emission from the hadronic phase is a relatively
small contribution and the measured photon spectrum directly
reflects the initial QGP phase of the evolution and hence the
initial temperature, this is not so at RHIC where the
QGP and hadronic contributions are found to be equally important.

In order to identify the reasons for this surprising behaviour, we
artificially switch off the transverse flow and recalculate the
photon spectrum. The result is shown in Fig.~\ref{F-1} (d).
This clearly demonstrates that
it is the comparatively strong flow in the late
hadronic evolution at RHIC (which is in this framework motivated by
the strong falloff of the HBT correlations with transverse momentum, see
\cite{RHIC-Model}) which leads to the observed importance of hadronic
contribution.

An additional factor is the different cooling in the hadronic phase:
the system created at SPS conditions is characterized by a large net
baryon density at midrapidity. This in turn implies more production
of heavy baryonic resonances at the phase transition at SPS as compared
to RHIC. Decay processes of these resonances in turn lead to an overpopulation
of pion phase space as compared to equilibrium, more so at SPS than at RHIC, 
which in turn causes more efficient cooling (see \cite{Comprehensive} and \cite{RHIC-Model}
for details). In essence, there is a longer-lasting hadronic phase at RHIC which
is hotter on average, leading to more photon emission and increasing the
relative contribution from the hadronic phase.

This is, as it stands, rather unfortunate since the slope of the measured
photon spectrum is no longer a good measure for the initial
temperature at RHIC but rather for the flow in the late phases, which
is comparatively well-known from the measured hadronic spectra and 
HBT correlations already. 

\section{HBT correlations for photons}

\begin{figure*}[htb]
\epsfig{file=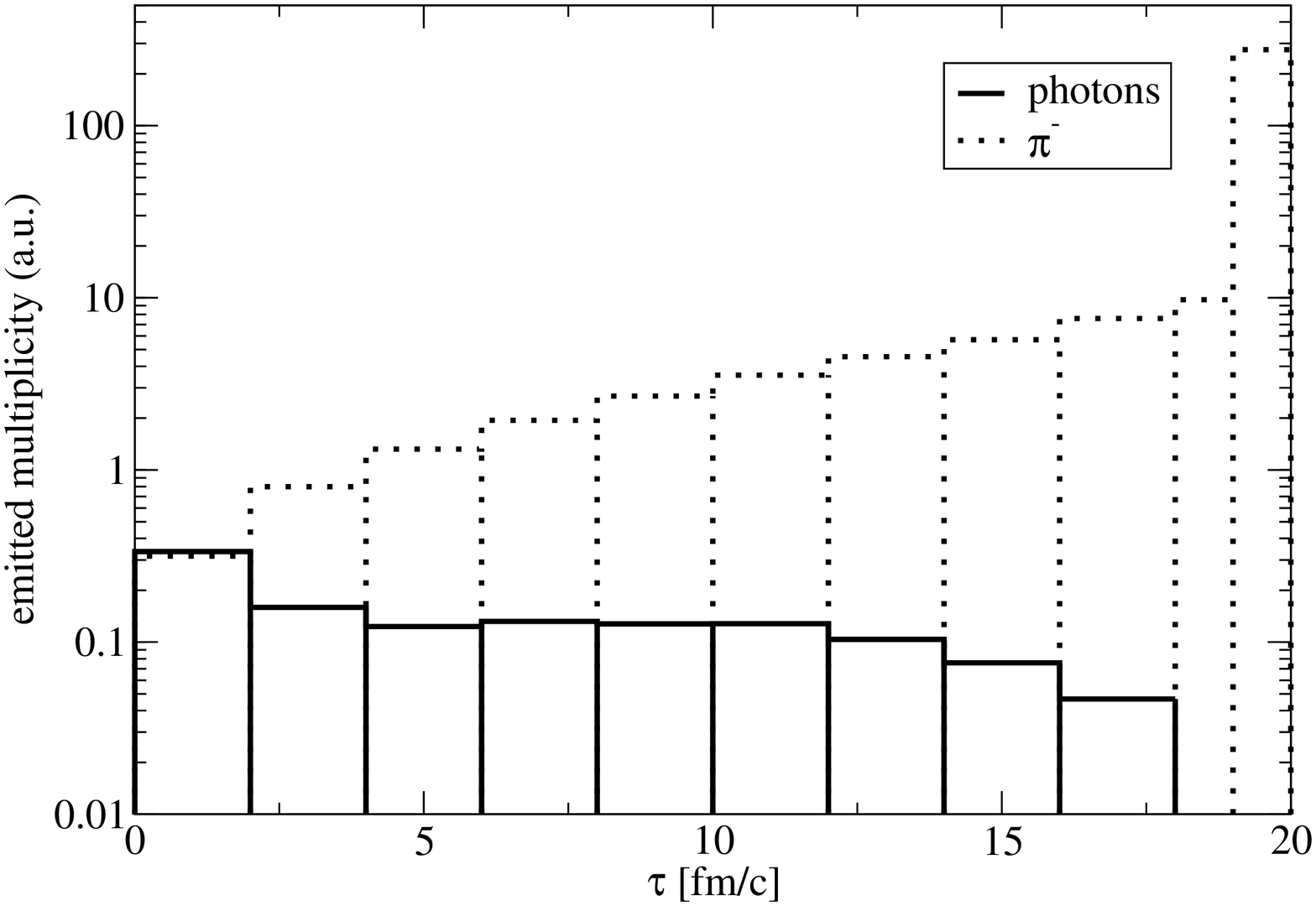, angle=-90, width=8cm}
\caption{\label{F-Times}The emitted multiplicity of photons (solid line) and
neg. pions (dotted line) at midrapidity in the momentum regions displayed in 
Figs.~\ref{F-RSide}, \ref{F-ROut} and \ref{F-RLong}
as a function of proper time $\tau$. Pion emission from
the fireball grows roughly with the surface area until finally the hot matter
decouples and the remaining pions are freed, therefore the emission is strongly peaked
towards late times. In contrast, photon emission takes place from the whole
volume at all times and is mainly a function of the average temperature at a given time, hence
there is a large contribution from early times.}
\end{figure*}

Using the folding of the rate with the fireball evolution
as emission function $S(x,K)$ (describing the amount of photons
with momentum $K^\mu$ emitted at spacetime point $x^\mu$) 
we calculate the HBT parameters as \cite{HBTReport, HBTBoris}
\begin{equation}
R_{\text{side}}^2 = \langle \tilde{y}^2 \rangle \quad R_{\text{out}}^2 =\langle (\tilde{x} -\beta_\perp \tilde{t})^2 
\rangle \quad R_{\text{long}} = \langle \tilde{z}^2 \rangle
\end{equation}
with $\tilde{x}_\mu = x_\mu -\langle x_\mu \rangle$ and
\begin{equation}
\langle f(x)\rangle(K) = \frac{\int d^4 x f(x) S(x,K)}{\int d^4x S(x,K)}
\end{equation}
Since photons escape without rescattering from all
spacetime points, we expect $R_\text{side}$ to be smaller than 
in a measurement of hadronic correlations, reflecting the fact that 
photons from the whole volume enter the emission function at all times (in contrast to
hadron emission which takes place from the surface until final breakup),
and we expect the difference of $R_\text{out}$ and $R_\text{side}$ to
be more pronounced, based on the observation that photon
emission is not dominated by a small time interval at the final
breakup of the fireball but shows a broader distribution (see Fig.¸\ref{F-Times}). 

\begin{figure*}[!htb]
\epsfig{file=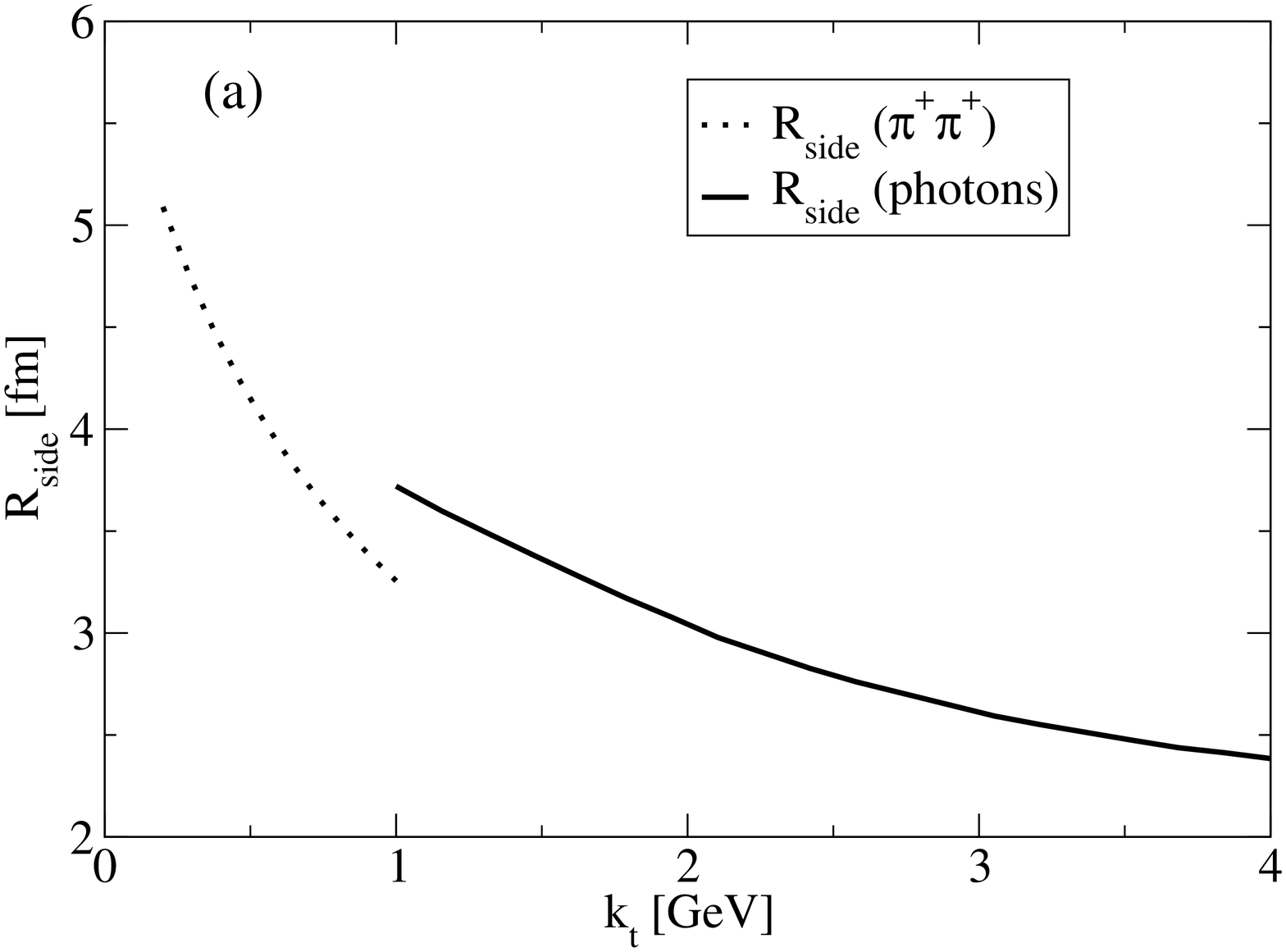, angle=-90, width=8cm}
\epsfig{file=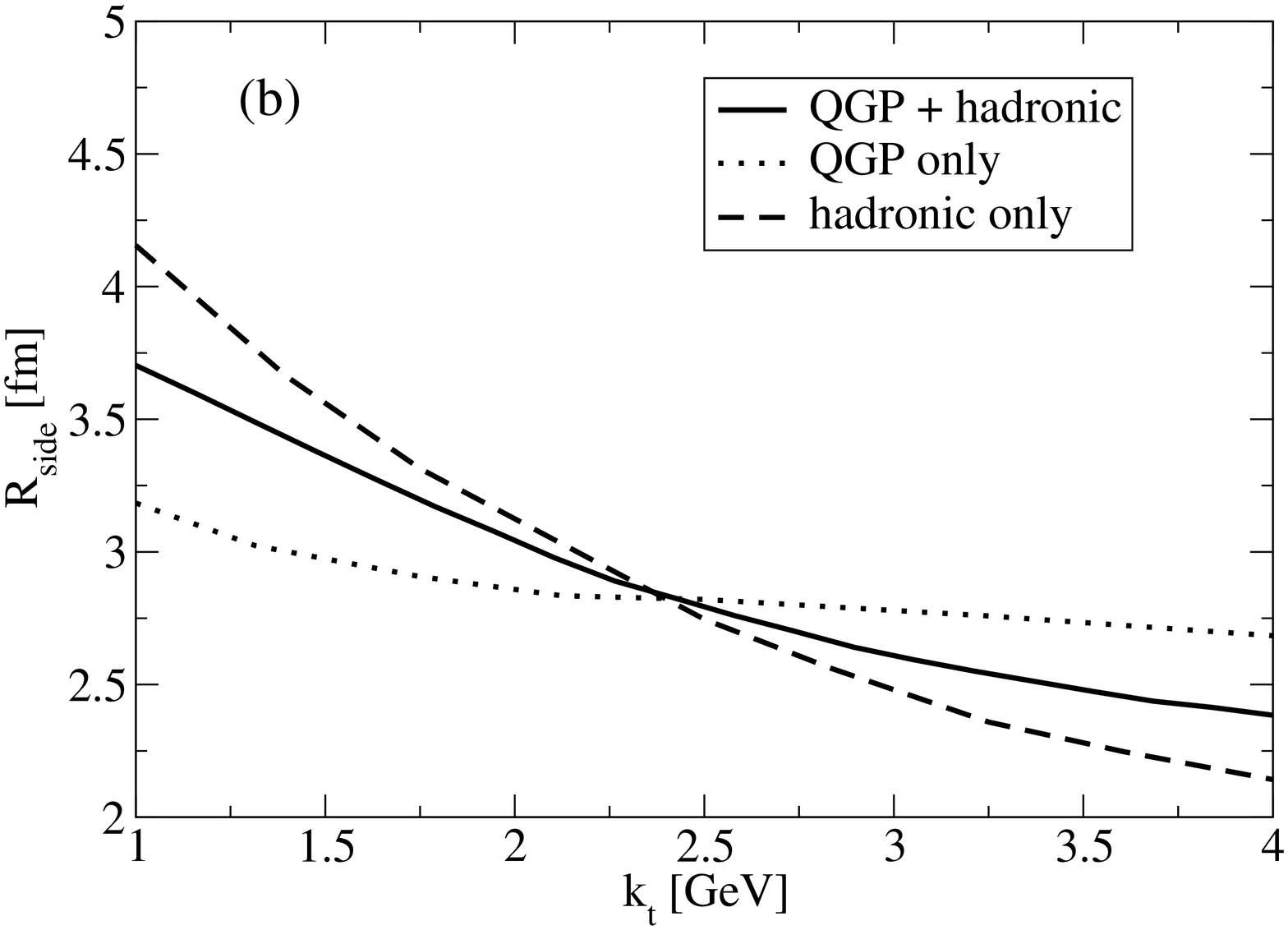, angle=-90, width=8cm}
\caption{\label{F-RSide}Left panel: $R_\text{side}$ as a function of transverse pair momentum
$k_t$ for photons (solid) and $\pi^+$ (dotted). Right panel: Decomposition
of the photonic $R_\text{side}$ (solid) into correlations among photons emitted from
the QGP (dotted) and hadron gas (dashed).}
\end{figure*}

\begin{figure*}[!htb]
\epsfig{file=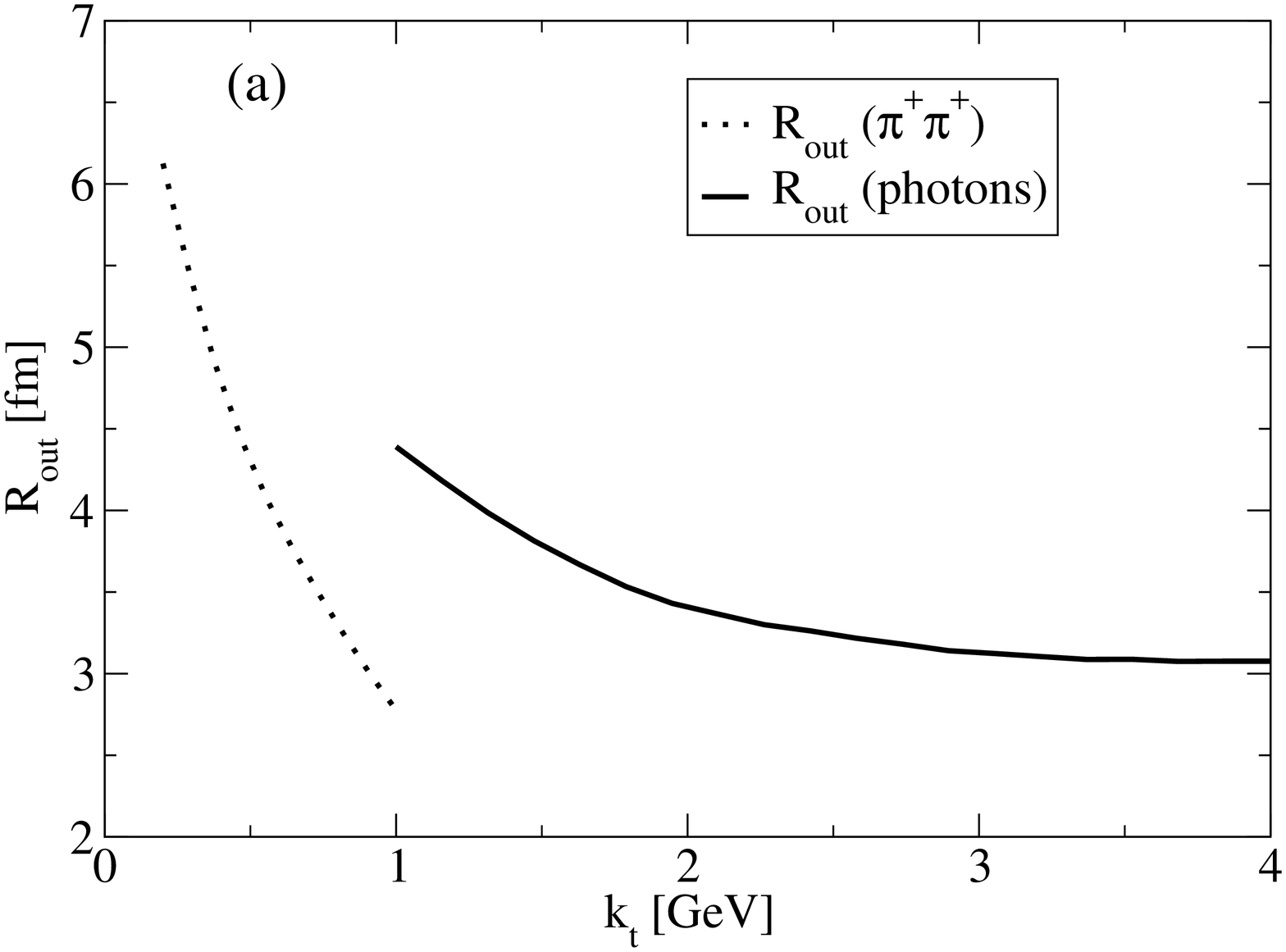, angle=-90, width=8cm}
\epsfig{file=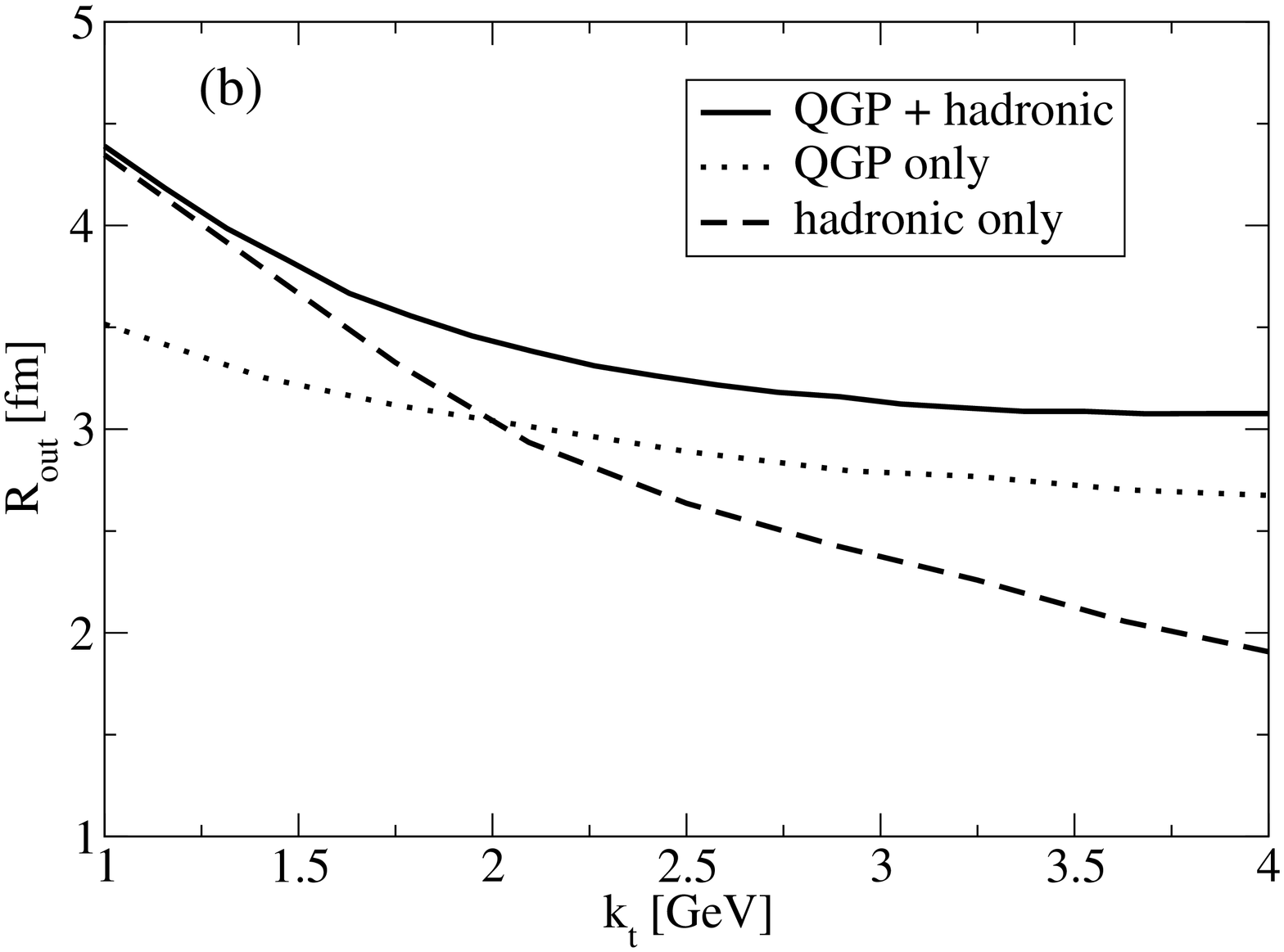, angle=-90, width=8cm}
\caption{\label{F-ROut}
$R_\text{out}$ as a function of transverse pair momentum
$k_t$ for photons (solid) and $\pi^+$ (dotted). Right panel: Decomposition
of the photonic $R_\text{out}$ (solid) into correlations among photons emitted from
the QGP (dotted) and hadron gas (dashed).
}
\end{figure*}

\begin{figure*}[htb]
\epsfig{file=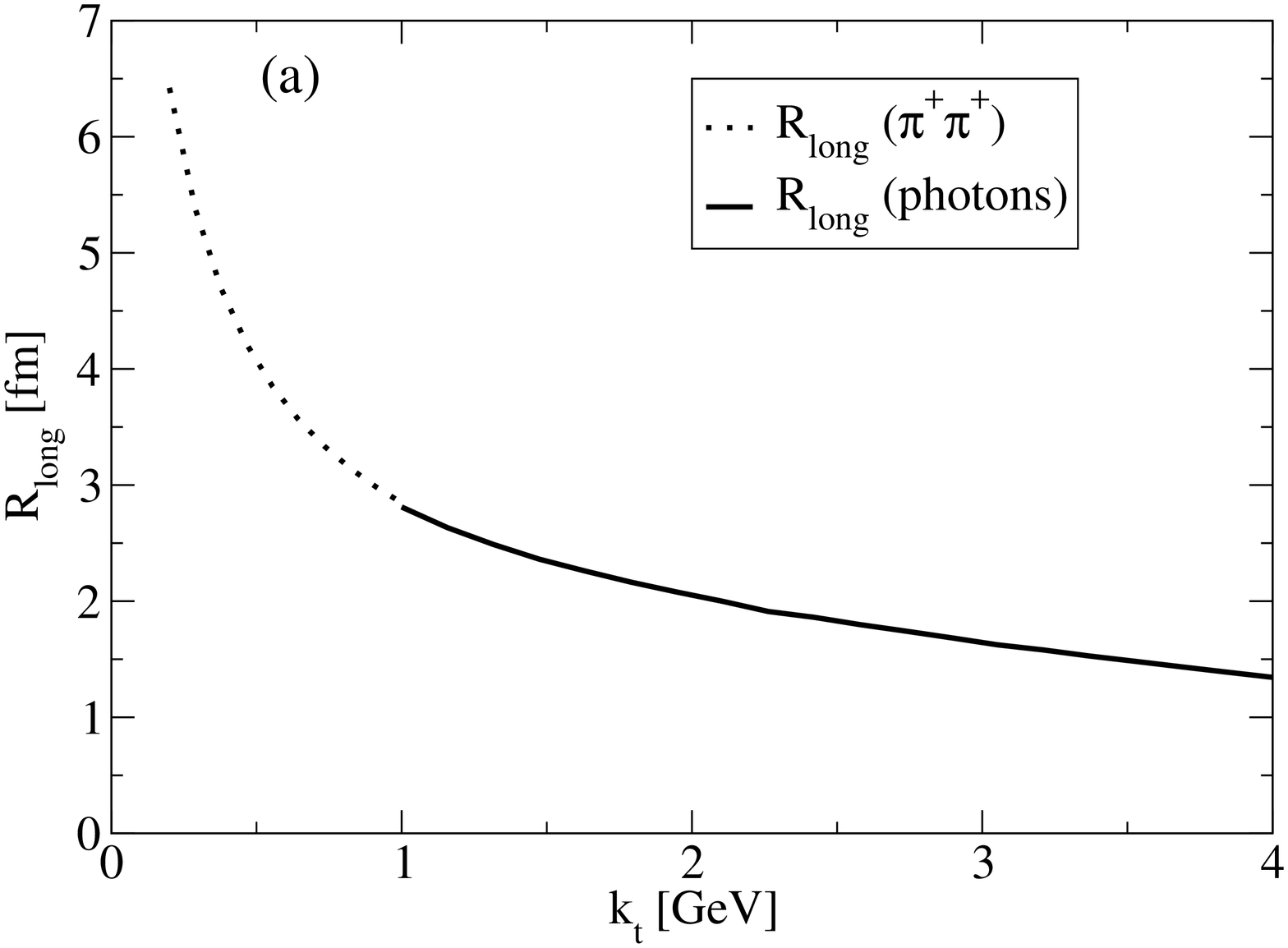, angle=-90, width=8cm}
\epsfig{file=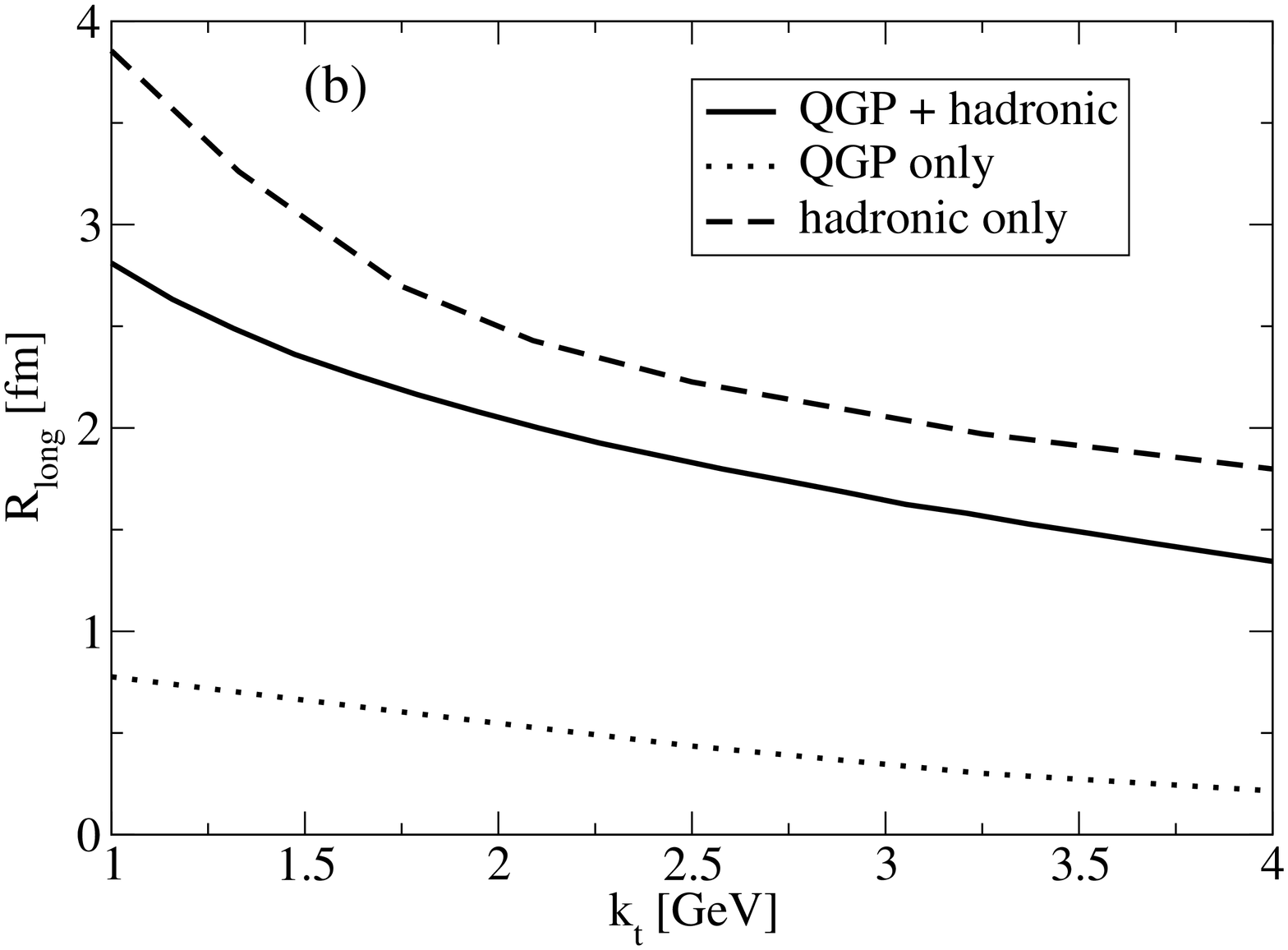, angle=-90, width=8cm}
\caption{\label{F-RLong}$R_\text{long}$ as a function of transverse pair momentum
$k_t$ for photons (solid) and $\pi^+$ (dotted). Right panel: Decomposition
of the photonic $R_\text{long}$ (solid) into correlations among photons emitted from
the QGP (dotted) and hadron gas (dashed).}
\end{figure*}

In Figs.\ref{F-RSide}, \ref{F-ROut} and 
\ref{F-RLong} (left panels) we show the three correlation radii and compare
with the hadronic HBT in the measured region \cite{RHIC-Model} (we do not make
any calculations for low  momentum photons since the emission rates
based on perturbative expansions cannot be expected to be valid in
this regime).

These expectations are to some degree confirmed. $R_\text{side}$ 
appears to be unexpectedly larger for photons than for pions,
but in order to deduce the geometrical size, one has to take into
account the effects of flow, and here the less steep falloff
of the photon $R_\text{side}$ as compared to the $\pi^+$ measurement
(reflecting the fact that a sizeable number of photons is emitted
early before transverse flow is built up) seems to extrapolate to a smaller value at 
$k_t =0$. There is a pronounced difference between $R_\text{out}$
and $R_\text{side}$ as expected before, and $R_\text{long}$ appears to be
approximately 
the same for photons and pions. This is not a surprise since 
the question of the transverse emission coordinate and hence the
difference between surface and volume emission is irrelevant here.

In order to gain more insight into the underlying physics, we present
the photonic correlation parameters as they would appear if one could
separate hadronic and QGP emission experimentally (Figs.\ref{F-RSide}, \ref{F-ROut} and 
\ref{F-RLong}, right panels). Consistently, the hadronic correlation
radii show larger absolute normalization and steeper falloff with $k_t$,
both features caused by the flow which at the same time expands the geometrical
radius and tends to descrease the size of the observed region of homogeneity
with $k_t$. 

In contrast, the early emission of QGP photons shows
a comparatively small-sized system characterized by little flow.

Note that $R_\text{side}$ for the total emission is to a good approximation an
average of the individual contributions weighted by their magnitude.
However, this does not hold for $R_\text{out}$ which includes a temporal component
as well: The sum of emission durations of hadronic and QGP phase has
to be larger than the individual emission durations.

\section{Comparison with other works}

Photon interferometry has been investigated in a number of studies
\cite{pInterferometry1, pInterferometry2,pInterferometry3,pInterferometry4,pInterferometry5}.
Here, we focus on a recent study \cite{Interferometry} based on the parton cascade
VNI/BMS \cite{VNI} and a subsequent boost-invarinat hydrodynamical evolution.

This particular model shows two pronounced differences to the present approach ---
is contains a pre-equilibrium contribution to the photon yield (not included here) 
and it is based on a boost-invariant expansion whereas our approach
is characterized by accelerated longitudinal expansion.

In order to make a meaningful comparison, we have to correct for the fact that
the correlation parameters in \cite{Interferometry} are calculated for
central collisions whereas in our approach we have chosen to calculate for
the 30\% most central collisions in order to be able to compare with the $\pi^+$
HBT results of the model (which describe in turn the measurements). This amounts
roughly to a 15\% increase of the radii when we go from the
values shown in Figs.\ref{F-RSide},\ref{F-ROut} and \ref{F-RLong} to central collisions.

In \cite{Interferometry}, $R_\text{side}$ at $k_t = 1$ GeV is found to be 4.6 fm,
at $k_t = 2$ GeV 3.8 fm. In our model we find for central collisions
similar (though slightly smaller) values of 4.3 fm and 3.5 fm 
respectively. Thus, $R_\text{side}$ for photons mainly reflects the initial 
spatial extension of the emission source and shows little signs of the
late source expansion.

The values for $R_\text{out}$ given in \cite{Interferometry} are 4.7 fm for $k_t = 1$ GeV
and 3.7 fm for $k_t = 2$ GeV, in contrast to 5.11 fm and 3.9 fm in the present model.
Presumably this is a consequence of the different distribution of emission in time:
Whereas in \cite{Interferometry} the emission at $k_t = 2$ GeV is almost completely
dominated by pre-equilibrium contributions with extremely short duration 
(leading to $R_\text{out} \approx R_\text{side}$) this
is different in our model, causing $R_\text{out} > R_\text{side}$.

$R_\text{long}$ again shows only a moderate difference between the two models: 
In \cite{Interferometry},
at $k_t=2$ GeV a value of 1.6 fm is found as compared to 1.95 fm in the present
model. In \cite{Interferometry}, this is a 
consequence of the dominance of pre-equilibrium
photons in this momentum region which are mainly emitted at times $\tau< 0.3$ fm/c, giving
little room for longitudial expansion, the observed value of $R_\text{long}$ is then
mainly dictated by the uncertainty principle. In contrast, 
the emission of thermal photons
in the present model starts only at 0.6 fm/c when the system has undergone some expansion.
Moreover, early longitudinal flow is stronger for a boost-invariant expansion, leading to
a further reduction of $R_\text{long}$ in \cite{Interferometry}. Given the fact that we do not
include a pre-equilibrium component of photon emission, our value of
2 fm probably overestimates the data somewhat. 

The effect of a pre-equilibrium contribution on the
HBT correlations depends on the relative magnitude of the contribution at given transverse
momentum. Currently, there is still considerable uncertainty about
the absolute magnitude of a pre-equilibrium contribution caused by 
questions like the role of LPM suppression. According to \cite{VNI-Photons} this 
may reduce the pre-equilibrium yield by up to a factor 3.5 at 2 GeV transverse
momentum. Presumably, a measurement of $R_\text{long}$ would be most capable
of answering the question if a pre-equilibrium contribution constitutes a 
sizeable fraction of the measured photon spectrum below 3 GeV or not.

\section{Summary}

We have presented predictions for the spectrum of thermal
photons at full RHIC energy and for two photon correlation
radii. This was done in a model which is able to describe both
the measured hadronic single-particle spectra and two-particle correlations.

We find that the photon spectrum between transverse momenta
2 GeV $< k_t < 4$ GeV is not dominated by emission from the hot QGP
phase as for SPS conditions \cite{Photons}. Instead, thermal
emission from the hadronic phase turnes out to be a significant
contribution. This is mainly due to the strone transverse flow
at RHIC (which in turn is required for a good description of the
hadronic HBT correlations).

Photon HBT correlations are expected to show a smaller system
characterized by less flow than seen in hadronic HBT measurements.
This is a consequence of surface versus volume emission processes.
However, in the present approach HBT correlations are not exclusively  dominated
by the QGP phase but reflect an average of the QGP and hadronic 
evolution phases. Given the importance of thermal photon radiation from
the hadronic phase, this is not surprising.

If this scenario turns out to be realized at RHIC then an extraction
of QGP properties from photon spectra and correlations may prove
difficult. Possibly the cleanest window would then be the high
momentum tail dominated by pre-equilibrium physics where
the number of emitted photons may still help to understand how
frequently scattering processes occur.

%\section*{Acknowledgements}

\begin{acknowledgments}

I would like to thank S.~A.~Bass and B.~M\"{u}ller for helpful discussions, comments and their
support during the preparation of this paper.
This work was supported by the DOE grant DE-FG02-96ER40945 and a Feodor
Lynen Fellowship of the Alexander von Humboldt Foundation.
\end{acknowledgments}

\end{document}